\documentclass[conference]{IEEEtran}

\usepackage{cite}  
\usepackage{siunitx}
\usepackage{xspace}
\usepackage{graphicx}  
\usepackage{multirow}
\usepackage{listings}
\usepackage{balance}

\usepackage{cite}
\usepackage{amsmath,amssymb,amsfonts}
\usepackage{algorithmic}
\usepackage{graphicx}
\usepackage{textcomp}
\usepackage{xcolor}
\usepackage[hidelinks]{hyperref}

\makeatletter
  \def\sectionautorefname{\S\@gobble}
  \def\subsectionautorefname{\S\@gobble}
  \def\subsubsectionautorefname{\S\@gobble}

\makeatother

\definecolor{mGreen}{rgb}{0,0.6,0}
\definecolor{mGray}{rgb}{0.5,0.5,0.5}
\definecolor{mPurple}{rgb}{0.58,0,0.82}
\definecolor{backgroundColour}{rgb}{0.95,0.95,0.92}
\lstdefinestyle{PythonStyle}{
	language=Python,
	backgroundcolor=\color{backgroundColour},
	commentstyle=\color{mGreen},
	keywordstyle=\color{mPurple},
	numberstyle=\tiny\color{mGray},
	stringstyle=\color{mGreen},
	basicstyle=\ttfamily\linespread{0.9}\scriptsize,
	breakatwhitespace=false,
	breaklines=true,
	captionpos=b,
	keepspaces=true,
	numbers=left,
	numbersep=5pt,
	showspaces=false,
	showstringspaces=false,
	showtabs=false,
	tabsize=2,
    frame=lrtb
}
\newif\ifdraft
\ifdraft
  \newcommand{\ian}[1]{{\textcolor{red}{ Ian: #1 }}}
  \newcommand{\logan}[1]{{\textcolor{blue}{ Logan: #1 }}}
  \newcommand{\ryan}[1]{{\textcolor{magenta}{ Ryan: #1 }}}
  \newcommand{\greg}[1]{{\textcolor{purple}{ Greg: #1 }}}
  \newcommand{\rajeev}[1]{{\textcolor{orange}{ Rajeev: #1 }}}
  \newcommand{\ganesh}[1]{{\textcolor{green}{ Ganesh: #1 }}}
  \newcommand{\kyle}[1]{{\textcolor{teal}{ Kyle: #1 }}}
  \newcommand{\yadu}[1]{{\textcolor{brown}{ Yadu: #1 }}}
  \newcommand{\nathaniel}[1]{{\textcolor{cyan}{ Nathaniel:  #1 }}}
\else
  \newcommand{\ian}[1]{}
  \newcommand{\logan}[1]{}
  \newcommand{\ryan}[1]{}
  \newcommand{\greg}[1]{}
  \newcommand{\rajeev}[1]{}
  \newcommand{\ganesh}[1]{}
  \newcommand{\kyle}[1]{}
  \newcommand{\yadu}[1]{}
  \newcommand{\nathaniel}[1]{}
\fi

\newcommand{\funcx}{{FuncX}}
\newcommand{\colmena}{{Colmena}}
\newcommand{\globus}{Globus}
\newcommand{\proxystore}{ProxyStore}

\begin{document}
\bstctlcite{IEEEexample:BSTcontrol}  

\title{Cloud Services Enable Efficient AI-Guided Simulation Workflows across Heterogeneous Resources}
\author{\IEEEauthorblockN{
    Logan Ward\IEEEauthorrefmark{1}, J. Gregory Pauloski\IEEEauthorrefmark{2}, 
    Valerie Hayot-Sasson\IEEEauthorrefmark{2}, Ryan Chard\IEEEauthorrefmark{1}, 
    Yadu Babuji\IEEEauthorrefmark{2}, Ganesh Sivaraman\IEEEauthorrefmark{1}, \\ Sutanay Choudhury\IEEEauthorrefmark{3},
    Kyle Chard\IEEEauthorrefmark{2}\IEEEauthorrefmark{1},   Rajeev Thakur\IEEEauthorrefmark{1}, and Ian Foster\IEEEauthorrefmark{1}\IEEEauthorrefmark{2}
}
\IEEEauthorblockA{
    \IEEEauthorrefmark{1}Argonne National Laboratory, Lemont, IL 60439;
    \IEEEauthorrefmark{2}University of Chicago, Chicago, IL 60637; \\
    \IEEEauthorrefmark{3}Pacific Northwest National Laboratory, Richland, WA 99354
}}



\maketitle
\thispagestyle{plain}
\pagestyle{plain}

\begin{abstract}
Applications that fuse machine learning and simulation can benefit from the use of multiple computing resources, with, for example, simulation codes running on highly parallel supercomputers and AI training and inference tasks on specialized accelerators.
Here, we present our experiences deploying two AI-guided simulation workflows across such heterogeneous systems. 
A unique aspect of our approach is our use of cloud-hosted management services to manage challenging aspects of cross-resource authentication and authorization, function-as-a-service (FaaS) function invocation, and data transfer.   
We show that these methods can 
achieve performance parity with systems that rely on direct connection between resources.
We achieve parity by integrating the FaaS system and data transfer capabilities with a system that passes data by reference among managers and workers, and a user-configurable steering algorithm to hide data transfer latencies.
We anticipate that this ease of use can enable routine use of  heterogeneous resources in computational science.

\end{abstract}


\section{Introduction}

Scientific applications that employ AI models often require or can benefit from the use of multiple, heterogeneous computing resources. 
For example, an application that uses AI-based perception to process data from a scientific instrument may engage a remote, more powerful computer for expensive tasks~\cite{vescovi2022linking},
while an AI-guided simulation application may run simulation codes on a conventional high performance computing (HPC) system but perform AI model training and inference on specialized accelerators~\cite{brace2021stream}.
While such multi-resource (or even multi-site) applications are far from new~\cite{smarr1992metacomputing,brunett1998application}, they are arguably becoming more commonplace due to more specialized computer architectures (especially for AI applications), increasingly portable code, and faster and more reliable networks.

The challenges around deploying a single application on multiple resources are well known.
Modern workflow systems typically place services on each resource (e.g., pilot jobs) that connect back to a controller through an open port on a central server or via a secure tunnel.
However, this approach requires either maintaining a server or establishing secure tunnels to each resource, introducing deployment complexity and single points of failure.
Hybrid software-as-a-service approaches, as used in Globus~\cite{chard2014globus} and \funcx{}~\cite{chard20funcx}, 
reduce deployment complexity and increase reliability
by using a cloud-hosted coordination service to manage authentication and respond appropriately to errors.

A second challenge in multi-resource deployments is to mitigate the costs of data transfer---a critical need in data-heavy AI applications.
Most multi-resource workflows communicate data between resources via the workflow controller or a shared data store like a database or shared file system.
Complexity arises, however, when data sizes increase and overwhelm the controller or resources differ in their access to shared data transfer mediums.
Passing data by reference and employing peer-to-peer data transfers can reduce strain on the workflow controller (e.g., Brace et al.\cite{brace2021stream}).
Separation allows data transfer via more efficient mechanisms (e.g., Parsl~\cite{babuji19parsl} uses Globus).
Subsystems that enable robust and flexible data movement by reference are key to the efficacy of multi-resource workflows.
%


We describe here our successes in deploying two multi-resource applications (molecular design and surrogate training) with approaches that maximize performance and minimize deployment challenges.
Both applications require using CPU resources to run simulation tasks and GPU resources to achieve timely solutions for AI tasks but vary in the frequency, duration, and data transfer requirements for each type of task. 
Our implementation uses cloud-managed services where possible to ensure reliability and simplify deployment of both secure function execution and data transfer between resources.
Specifically, we chose a federated FaaS platform (\funcx{}) and a system (\proxystore{}) that enables peer-to-peer data transfer via \globus{} with minimal changes to application code.
We first characterize the performance of these tools using a synthetic application and then describe how we achieved performance parity between our cloud-managed solution and one using a conventional workflow system, Parsl.

\section{Related Work}\label{sec:background}
Many capabilities are needed to implement multi-resource computational campaigns.
Specifically, workflow systems must
1) execute tasks on resources acquired from multiple facilities; 
2) exchange large task data between resources;
3) express policies that avoid latencies inherent to orchestrating multiple systems.
Here, we review the technologies available for each need.

\subsection{Coordinating Remote Execution}

Remote computing has often been performed via SSH connections. 
Grid computing introduced remote connectivity and web services to manage remote execution, for example via Globus Grid Resource Allocation Manager (GRAM)~\cite{feller2007gt4}. Recently, facilities such as NERSC and TACC have developed web APIs for submitting and managing batch jobs with Newt~\cite{cholia10newt} and Tapis~\cite{stubbs21tapis}, respectively. These approaches 
add each new task to a global queue, which can result in significant delays due to the need to wait for batch jobs to be scheduled in applications where new tasks are created dynamically.

Many workflow systems have been developed to execute sets of tasks on both local and remote computing resources. 
Systems such as Pegasus~\cite{deelman19pegasus}, Dask~\cite{rocklin2015dask}, Parsl~\cite{babuji19parsl}, Swift~\cite{zhao07swift}, and 
Galaxy~\cite{afgan22galaxy} differ in various ways, but all provide:  
1) a method for describing a workflow;
2) a data model for representing dependencies among tasks (e.g., a DAG); and
3) a runtime system for executing tasks on local and/or remote resources.
When dispatching tasks to HPC, 
such systems often employ multi-level scheduling schemes that map individual tasks to workers deployed on HPC resources, an approach that 
allows dynamically generated tasks
timely access to computing resources~\cite{bala19radical}.



The serverless or FaaS paradigm popularized by hosted cloud platforms such as Amazon Lambda~\cite{amazonlambda},
Google Cloud Functions~\cite{googlecloudfunctions}, and Azure Functions~\cite{azureFunctions} is an attractive model for remote scientific computing as it 
allows large applications to be broken down into smaller, efficient, 
and adaptable functions~\cite{foster2017cloud,spillner2017faaster,malawski2016towards,fox2017conceptualizing,kiar2019serverless}. 
However, the commercial FaaS offerings are proprietary systems
that are restricted to a single cloud provider and do not support
execution of functions on existing cyberinfrastructure. 

Open source frameworks (e.g., Apache OpenWhisk~\cite{openwhisk}, Fn~\cite{Fn}, KNIX MicroFunctions~\cite{knix},  Abaco~\cite{garcia2020abaco}) can be used for on-premise deployments.
However, most 
use Docker and rely on Kubernetes to operate---an assumption that prohibits
use on HPC systems that employ batch schedulers.
Some systems, such as ChainFaaS~\cite{ghaemi2020chainfaas} and DFaaS~\cite{ciavotta2021dfaas}, support distributed function execution on personal computers and edge nodes. 
\funcx{}~\cite{chard20funcx} is the only system that supports remote execution on a federated ecosystem of endpoints in a diverse research cyberinfrastructure spanning from HPC to edge.

\subsection{Inter-Resource Data Fabrics}

Workflow systems that couple diverse applications in distributed environments require a data fabric which provides consistent access to data regardless of location.
In the tuple space model, originating in Linda~\cite{carriero1994linda}, data producers and consumers use put and get operations on a shared distributed-memory space.
Dataspaces~\cite{dataspaces2017aktas} supports large-scale, dynamic scientific applications via a tuple-space-like model implemented using the Margo and Mercury RPC libraries~\cite{soumagne2013mercury, ross2020mochi}.
WA-Dataspaces~\cite{aktas2017wa} provides predictive prefetching and data staging support in wide area networks.

Multi-site workflows require a secure data fabric accessible by hosts behind different firewalls, features not supported by Linda, Dataspaces, or WA-Dataspaces.
SSH tunnels can enable secure communication but can be cumbersome to establish and fragile to maintain.
Science DMZs~\cite{data2013sciencedmz} provide secure sub-networks that span sites without firewalls but are only available at select computing sites.
SciStream~\cite{chung2022scistream} uses gateway nodes provided by computing sites for fast memory-to-memory data streaming for high throughput science applications.
Cloud services provide higher availability than the aforementioned but add additional network hops (which adds latency), fail to take advantage of high-performance connections between sites, and can be cost prohibitive.

\subsection{AI-Integrated Workflows}
AI-integrated workflows are emerging as a powerful tool across many scientific domains, with many use patterns~\cite{jha2022AIHPC, ejarque2022aiworkflows}.
Following the lexicon of Jha et al.~\cite{jha2022AIHPC}, there are at least ``ML-in-HPC'' applications where AI-based software are used inside conventional HPC applications (e.g., surrogates for expensive computational routines~\cite{atlas2022AtlFast3}) and ``ML-out-HPC'' where machine learning controls the execution of the application (e.g., steering a workflow via active learning~\cite{dunn2019rocketsled, montoya2020camd, lee2019deepdrivemd}).
The large diversity of application types is reflected in the many tools that support integration of AI in existing scientific codes or the creation of entirely new classes of applications.
Most relevant to our work are variants that use AI to determine the next task to execute in a workflow.

Systems for steering AI-guided workflows vary in how they express coordination between the workflow and the intelligence system.
Systems such as Supervisor~\cite{wozniak2018supervisor} and DeepHyper~\cite{balaprakash2018deephyper}
use a model where a single process running a steering algorithm receives results and submits new tasks to the workflow via a queue. 
In libEnsemble~\cite{libEnsemble}, a workflow is expressed as \textit{simulation} tasks that produce data, 
\textit{generator} tasks that produce new tasks,
and \textit{allocator} tasks that determine when to launch each type of task.
Ray~\cite{mortiz2018ray} and Decaf~\cite{yildiz2021decaf} allow a decentralized model where the steering logic is expressed in an agent-based programming model.


\section{Motivating Applications}

\begin{figure}
    \centering
    \includegraphics[trim=1mm 0mm 1mm 0mm,clip]{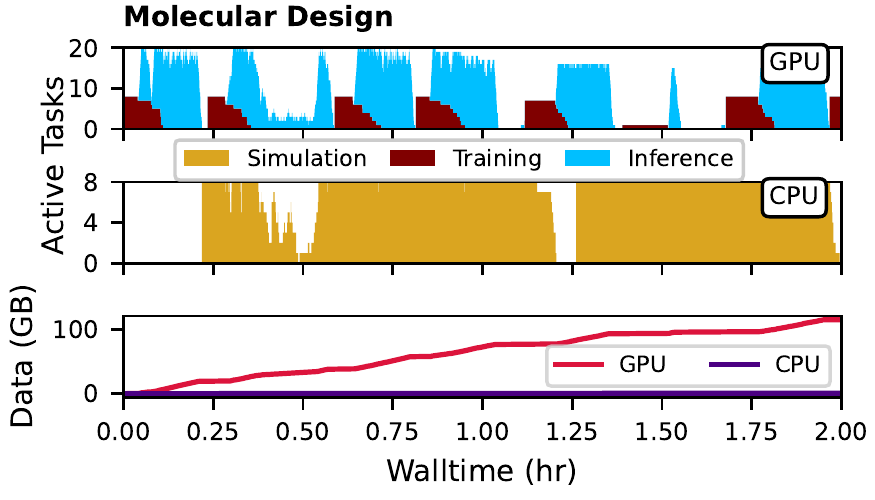}
    \includegraphics[trim=1mm 0mm 1mm 0mm,clip]{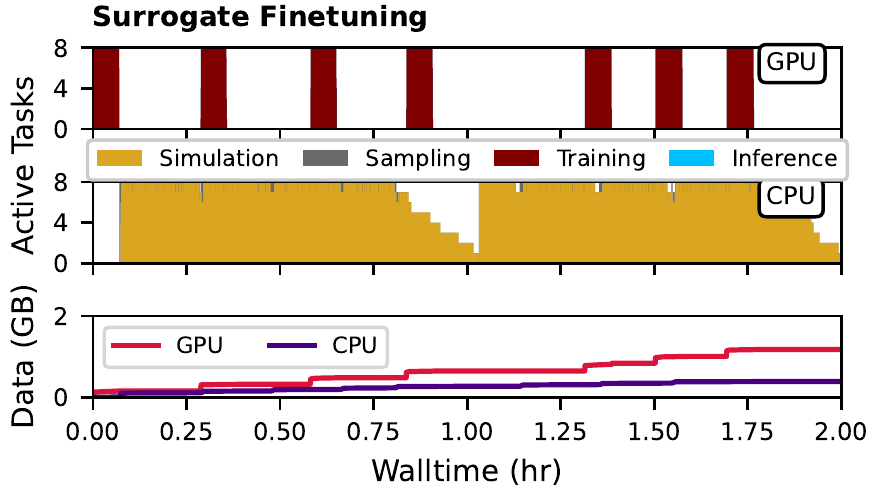}
    \caption{Resource utilization for our example applications. We show the number of tasks running and the cumulative data transfered to each resource over time. Data were collected using 20 T4 GPUs and 8 KNL processors, with a workflow based on Parsl without pass-by-reference.}
    \label{fig:apps}
\end{figure}

We study two applications that require CPU resources for simulation tasks and GPU resources for AI tasks. 
Each have different coordination patterns and data transfer requirements, as shown in \autoref{fig:apps}.
The first application, Molecular Design, consistently runs tasks on a GPU and transfers O(10)~GB
per batch of AI tasks.
The second application, Surrogate Fine-tuning, runs AI tasks more sporadically and has an order of magnitude smaller data transfer requirements.
The quality of results in both applications requires completing AI tasks rapidly after sufficient simulation tasks have completed.

Full implementations are available on GitHub (\url{https://github.com/exalearn/multi-site-campaigns}) and
the Materials Data Facility~\cite{ward2023data}.

\subsection{Application: Molecular~Design}\label{sec:moldesign}

This application seeks to identify molecules within a candidate set (here, \num{1115321} molecules from the MOSES dataset~\cite{polykovskiy2020moses}, as collected in the nCov-Group Data Repository~\cite{babuji2020covdata}) that have
desirable properties.
Properties of any individual molecule can be determined via an expensive quantum chemistry simulation, so
we use active learning~\cite{cohn1996active} to select simulations to perform.
Results from previous simulations 
are used to train a machine-learned surrogate model, which is then used to infer scores for the remaining molecules; those scores are used to determine which simulations to be performed.
This process repeats until our computational budget is exhausted.
The application thus performs \textit{simulation}, \textit{training}, and \textit{inference} tasks.



\textbf{Simulation} The simulation task computes a molecule's ionization potential (IP), a key design property for organic electrolytes~\cite{li2020recent}, using tight binding, an inexpensive quantum-mechanical simulation method.
First, RDKit~\cite{landrum2006rdkit} is used to generate an initial 3D geometry for a molecule from its bonding connectivity.
Then, geoMETRIC~\cite{leeping2016geometric} determines the equilibrium geometry/energy for the neutral and charged states using the energies and forces computed with the QCEngine interface~\cite{smith2020qcarchive} of eXtended Tight Binding (xTB)~\cite{bannwarth2020xtb}.
All computations take $\sim$60~s on a CPU and produce 1~MB data.

\textbf{Training} The surrogate model used to predict a molecule's IP given its bonding connectivity comprises an ensemble of eight message-passing neural networks~\cite{stjohn2019mpnnpolymers} (MPNNs).
Each model has the same architecture but is trained on a different, randomly-selected subset of the training data.
Training each model requires 340~s on a GPU and generates 10~MB, and models can be trained in parallel.

\textbf{Inference} The inference tasks use the ensemble of MPNN models from the training task to estimate IPs.
After predicting all molecule IPs with each model,
the molecules are ranked by the Upper Confidence Bound (UCB) of the predictions, which is the sum of the mean and standard deviations of the model predictions.
Scoring the full dataset per model takes 900~s on a single GPU (1000~inferences/s) and requires transfer of 2.4~GB (model weights, molecule inputs, outputs). 


Success is measured by how many molecules we find with high IPs after a certain amount of compute has been expended.

\subsection{Application: Surrogate Fine-tuning}\label{sec:hydronet}


This application trains a surrogate model for expensive quantum mechanical computations.
Our goal is to produce a model capable of replicating the energies and forces from Density Functional Theory (DFT) calculations on clusters of water surrounding a methane solute.
We first train a model on the energies of many water clusters computed using an approximate method (Thole-Type Models~\cite{fanourgakis2006ttm}, HydroNet~\cite{choudhury2020hydronet} Dataset) and then refine the model with a small number of energies and forces of solvated methane calculated using DFT.
We use an active learning approach similar to the molecular design application to guide the choice of training data: we run the DFT calculations on the structures where the surrogate model is least certain.
Beyond the inference, training, and simulation tasks required by active learning, we also use a \textit{sampling} task to produce new structures.

\textbf{Training}: We train surrogate models using the SchNet architecture~\cite{schutt2018schnet} as implemented in PyTorch-Geometric.
A training task requires an average of 4~minutes on a GPU and transmitting 21~MB.
We train an ensemble of 8 SchNet models where each is trained on a different, randomly-selected subset of the training data.

\textbf{Inference}: An inference task on a batch of 100 structures takes an average of 3.2~s on a GPU and involves transmitting 3~MB  (including all inputs and outputs).

\textbf{Simulation}: Psi4~\cite{smith2020psi4} via the Atomic Simulation Environment (ASE) interface~\cite{larsen2017ase} is used to compute the energy and forces acting on a cluster of atoms using the PBE0 exchange-correlation function and the aug-cc-pvdz basis set.
Each task takes a mean of 360~s on CPU and produces 20~kB.

\textbf{Sampling} Sampling tasks produce new structures using a trained SchNet model.
We create new structures by initializing the temperature of a structure of water-solvated methane to 100K, then running molecular dynamics for a set number of timesteps.
Choosing the number of timesteps involves a tradeoff: too few produces too little diversity in structures, 
too many produces unrealistic structures due the accumulation of model errors over many steps.
Our first sampling tasks use only 20 timesteps because our model is not yet capable of producing realistic dynamics.
We gradually increase to 1000 over the course of the run.
Sampling tasks take between 1 and 3~s on a CPU and require transmitting 3~MB.


Our steering agent selects the next simulation to perform from two pools, 
each designed to provide unique structures to add to the training set.
The \textit{audit} pool contains the last structure from each sampling task, which should be the most different from the training set by nature of being farthest along a time-evolution pathway.
The \textit{uncertainity} pool contains the structures for which our models have the least-certain predictions across all structures produced during sampling.
We re-populate the uncertainty pool each time 100 new structures have been sampled by performing inference on each newly-sampled structure with each model in our ensemble and ranking structures based on the variance in predicted energy.

We start by training our model ensemble on \num{1720} previously collected structures and continue the active learning until 500 new structures have been added to the set.
Our steering algorithm balances the number of workers devoted to simulation and sampling to maintain a constant number of structures in the audit pool.
The algorithm begins a new training run after 25 new structures are added to the training set.

Success is measured by the accuracy of a model trained on all DFT calculations available after a run using test sets created from data unseen during training.
We produced the test set by running molecular dynamics for 10 structures of solvate methane using DFT to compute energies and forces at starting temperatures of 100K, 300K, and 900K for 32 timesteps.

\section{Technical Approach}

\begin{figure}
    \centering
    \includegraphics[width=\columnwidth]{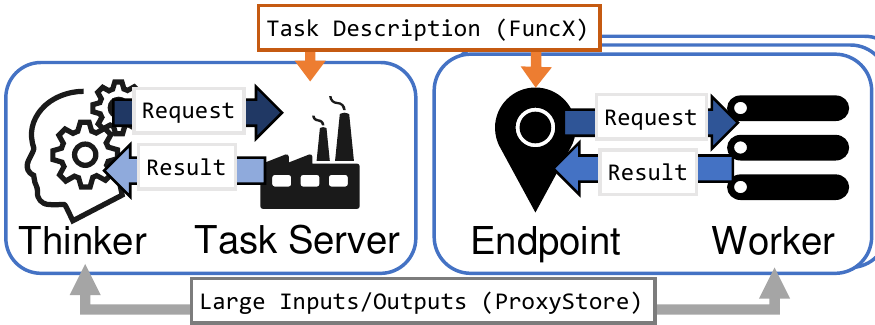}
    \caption{Our implementation of a multi-site workflow.
    A \textbf{Thinker} expressed using \colmena{} controls when tasks are executed by communicating with a \textbf{Task Server} that manages submitting task via \funcx{}. \funcx{} routes task requests through a cloud provider and uses endpoints deployed on multiple resources that communicate with workers deployed on compute nodes. Large data are passed directly between the Thinker and workers using \proxystore{} backed by \globus.}
    \label{fig:overall}
\end{figure}

As detailed in \S\ref{sec:background}, AI-guided applications require systems for remote execution, data transfer, and defining our coordination policy.
We selected tools to achieve several goals: security, performance and scalability, minimial network configuration, robustness, and portability.
These objectives lead us to use \funcx{} to provide secure and scalable
remote execution across resources, 
\proxystore{} to offload data transfer to different data fabrics (e.g., Globus),
and \colmena{} to express AI-guided workflows.
We describe each component below.

\subsection{Cloud-hosted management services}
Our approach centers around the use of cloud-hosted services to manage data and compute across distributed resources. 
This approach enables our applications to communicate with the reliable and accessible cloud-hosted Globus Transfer and \funcx{} services irrespective of data or compute location. 

\subsubsection{Security}
We require a security model that ensures only authenticated and authorized users can execute tasks and access data.
A multi-resource or -site workflow requires a solution that supports different identity management systems and authentication models, stringent security requirements (e.g., two-factor authentication, short authorization lifetimes), the need to use different accounts to access different resources, and resource sharing among groups of users. 

Globus and \funcx{} fulfill such security requirements through Globus Auth~\cite{tuecke2016globus}, a general identity and access management platform. Globus Auth is widely adopted in the scientific community, implements standard protocols (e.g., OAuth~2), and enables secure delegation so 
workflows (e.g., Colmena) can securely leverage other 
services like \funcx{} and Globus. 

\subsubsection{Network simplicity}
Both Globus and \funcx{} implement a hybrid deployment model, relying on users or administrators to deploy lightweight
software (Globus Connect or \funcx{} Agents) on remote resources. 
Communication between cloud and endpoints
use inbound HTTPS for Globus and outbound RabbitMQ TCP sockets for \funcx{}.
These endpoints are 
securely paired with the cloud platforms and
subsequent access is permitted only to authenticated and authorized users. 
Network connections between resources are not required.

\subsubsection{Robustness}
Cloud-hosted management services are highly available and provide high levels of robustness compared to locally-managed services.
For example, both \funcx{} and Globus's services accept and store tasks (and results) even while remote endpoints (or clients) are unavailable so tasks can be resumed when endpoints reconnect to the cloud.

\subsection{\funcx: Federated Function-as-a-Service} 


\funcx{} is a federated FaaS platform that combines
a hosted cloud service for registering Python functions and dispatching function invocations
with an ecosystem of user-managed endpoints deployed on arbitrary computing
systems.
Users can invoke functions by passing the function
body and input arguments to the cloud 
service via an implementation of Python's \texttt{concurrent.futures.Executor} interface.  
Unlike traditional FaaS systems, a \funcx{} user also supplies an endpoint ID that determines where 
the function will be executed. 
The \funcx{} endpoint is user-deployed and is responsible for provisioning
resources and managing function execution. The endpoint code can interface
with different schedulers (e.g., Slurm or PBS) to provision resources.
\funcx{} manages serializing
the function body and arguments, sending them to the remote
endpoint, and execution on the remote endpoint. 



\subsection{\proxystore} 
\label{sec:proxystore-backends}

Multi-resource workflows require a resource-spanning data fabric for data exchange.
Architecting this system element is challenging because capabilities and performance characteristics vary across resources and a single workflow may have many distinct data patterns (e.g., both frequent small transfers and infrequent large transfers).
Any data transfer mechanism provided by the compute fabric is typically designed to support the most common use case: e.g.,
\funcx{} communicates function inputs and results via the cloud which adds overheads and incurs financial costs, so that payloads are limited to 10~MB.
Decoupling the data fabric from compute fabric can enable greater flexibility in data management. 
Here, we use \proxystore{}, a system that enables pass-by-reference functionality in a diverse range of distributed applications without any modifications to application code.

Passing data by reference is key to dissociating data transfer from control flow. 
References are small so can be efficiently moved along with function bodies, and
can be passed between many resources without having to incur costs for the movement of the actual data.
That is, the data are only de-referenced once on the target resource regardless of how many intermediate resources through which the reference was passed.
This ensures processes or services responsible for control flow do not become an I/O bottleneck.

Rather than exposing traditional \texttt{get}/\texttt{set} operations, \proxystore{} uses the \emph{proxy} model for seamless pass-by-reference functionality in Python.
This easy-to-use programming paradigm allows users to dynamically change transfer methods without needing to modify task code.
ProxyStore supports many backend object stores and transfer methods to provide efficient transfer of objects between processes located within the same resource or across different resources.



Proxies are used to intercept and redefine operations on a \emph{target} object. 
\proxystore{} uses lazy transparent object proxies that behave identically to the target object which is achieved by the proxy forwarding all operations on itself to the target.
Lazy proxies provide just-in-time \emph{resolution} of the target via a \emph{factory} function.
Factories resolve the target when called; a proxy, initialized with a factory, calls the factory and retrieves the target the first time the proxy is accessed.

When the \texttt{proxy()} function is called on a target object, the object is placed in a backend store, a factory
capable of resolving the object from the store is created, and a proxy, initialized with the factory, is returned.
The resulting proxy is the lightweight reference that can be efficiently transmitted.
A function that receives a proxy uses the proxy as if it were the target object due to the proxy's transparent nature so no modification of function code is needed.

In this work, we use the Redis, file system, and \globus{} backends for \proxystore{}.
The Redis backend is ideal when resources exist within a single, fast network.
The file system backend supports scenarios where separate systems have access to a
shared file system. 
The \globus{} backend is used for multi-resource applications that lack a shared file system.

\subsection{\colmena: Steering Policies as Cooperative Agents}

\colmena{} is a Python library for expressing the steering of dynamic workflows as a collection of interacting agents, which are known collectively as a \textit{Thinker}. 
The Thinker controls what tasks are performed and how resources are allocated by a workflow system over time based on behavior defined in the agents.
For example, one agent may submit a retraining task after a certain number of simulation tasks complete, and another may submit a new simulation when resources are available.
Agents, each running as a separate Python thread, interact via Python's threading primitives (e.g., the simulation agent may consume tasks from a queue populated from a task scoring agent).
A Thinker communicates task requests to a \textit{Task Server} that employs some compute fabric (e.g., Parsl or \funcx{}) to execute tasks on distributed resources and return results back to the Thinker (\autoref{fig:overall}).

\colmena{} integrates support for \proxystore{} by automatically creating proxies for objects larger than a user-specified size.
The threshold size and \proxystore{} backend can vary between tasks types,
and users can also proxy objects manually before submitting the proxies to tasks.
The flexibility in how proxies are generated makes it possible to adopt different data fabrics between sites and to cache data needed for a computation ahead of time.
Regardless of whether \colmena{}'s automatic proxying is used or the user manually proxies objects, no changes to user task code or the compute fabric (i.e., \funcx{}) are necessary with \proxystore{}.

\section{Results and Discussion}
\label{sec:results}

We started by evaluating the data fabric in a simplified, synthetic application (\autoref{results:proxystore}),
then evaluated our cloud-managed approach in depth for one of the two applications (\autoref{results:design}).
Finally, we contrast the performance differences between a cloud-hosted approach and a baseline that lacks our advanced data fabric using both motivating applications (\autoref{results:compare}).

\subsection{Computational Resources and Software Configuration}

We use the computational resources of the Argonne Leadership Computing Facility (ALCF) and Computing, Environment, and Life Sciences (CELS) directorate for our experiments.
The Knights Landing nodes of ALCF's Theta supercomputer are used to perform the simulation components of each workflow.
We use a NVIDIA DGX server with 20 T4 GPUs (known as Venti) for AI tasks (i.e., training, inference).
The Venti system is representative of off-site resources because, even though housed in the same building, it exists on a separate network, does not have access to any Theta file systems, and uses a different authentication procedure.

The \colmena{} Thinker and Task Server reside on a login node of Theta.

\subsection{Workflow Configurations}\label{sec:workflow}

We compare three different workflow  system configurations. 
Our two baselines use Parsl, which requires open ports or a tunnel to each computing resource to communicate task information.
All configurations require creating a Python virtual environment on each computing resource.

\begin{enumerate}
    \item \textbf{Parsl}: Our baseline without \proxystore{}. Requires two ports for Parsl to communicate tasks to, and results from, a remote system.
    \item \textbf{Parsl+Redis}: Our baseline with \proxystore{} using Redis to communicate task data across sites and the file system for local tasks. Requires a third port for Redis in addition to two for Parsl.
    \item \textbf{FuncX+Globus}: Uses FuncX to communicate task instructions, and \proxystore{} with Globus for task data across sites and the file system for local tasks. Requires no open ports besides those used by Globus, which are already configured at most computing centers.
\end{enumerate}


\subsection{Evaluating Data Transfer using Synthetic Applications}\label{results:proxystore}

We first investigate using pass-by-reference to reduce overheads in \colmena{} and \funcx{}. Then we profile \proxystore{} backends to guide our deployment strategies. 

\begin{figure}
    \centering
    \includegraphics[trim=1mm 0.25cm 1mm 0, clip,width=\columnwidth]{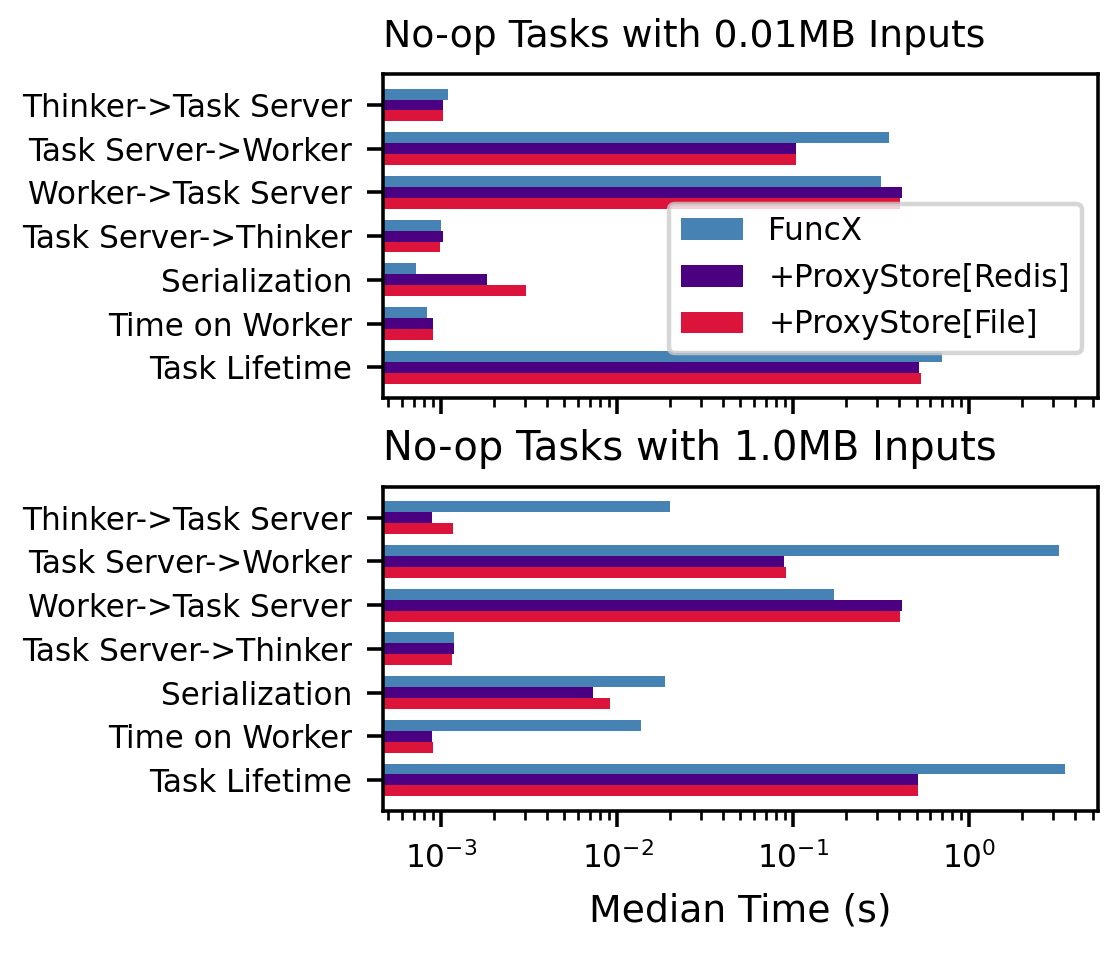}
    \caption{
    Median times for different components of the end-to-end execution of a no-op task with \colmena{}.
    The use of \proxystore{} to transfer objects reduces communication costs for both small (10~kB) and large (1~MB) task inputs.
    \proxystore{} can avoid repeated serializations and deserializations of object transmitted through the Task Server and \funcx{} service.
    }
    \label{fig:proxystore-overhead}
\end{figure}

\subsubsection{Avoiding \funcx{} Overheads with \proxystore{}}

We first compare the performance achieved when communicating task inputs with \funcx{} and two \proxystore{} backends: shared file system and Redis.
Specifically, we aim to quantify overhead improvements to the Task Server and \funcx{} by transferring large objects via alternative means through \proxystore{}.

We execute no-op tasks that return no output to measure task overheads. We perform the experiment with 10~kB and 1~MB inputs. We chose the input sizes  based on characteristics of \funcx{}.
\funcx{} stores function arguments and results smaller than 20~kB in an Amazon ElastiCache Redis store and objects greater than 20~kB in Amazon S3. 
In all experiments, the Thinker and Task Server are located on a Theta login node, and we use a \funcx{} endpoint on Theta which executes tasks on a single Theta KNL node.
We execute 50 tasks and record the time spent in different stages of the task's life cycle.

We show in \autoref{fig:proxystore-overhead} communication times between the Thinker, Task Server, and worker, as well as the serialization time, time spent on the worker, and overall task lifetime.
\emph{Serialization} time is that spent serializing and deserializing tasks on the Thinker, Task Server, and worker.
When serializing a task, \colmena{} scans for task inputs or outputs with sizes exceeding the \proxystore{} threshold (set to zero for this experiment).
If such large objects are found, the object is proxied and the lightweight proxy is serialized along with the task instead.
Therefore, the serialization time reflects proxying time, which includes time spent communicating objects to Redis or writing objects to disk.
\emph{Time on worker} is the time between the task starting on the worker and the worker returning the completed task;
it includes deserialization of the task, possible resolving of proxies, the execution of the task itself (which in this case is a no-op), and the serialization of the results.
\emph{Task lifetime} is the time between a task being created by the Thinker and the result being received by the Thinker.

Task Server-to-worker communication dominates the overall task lifetime because inputs must go through \funcx{}'s cloud service.
Passing objects via proxies can reduce this cost by 2--3$\times$ for 10~kB inputs and up to 10$\times$ for 1~MB inputs.

Similar magnitude speedups are found for the communication between the Thinker and Task Server with larger objects.
The Thinker and Task Server communicate via Redis queues so sending small objects (e.g., 10~kB) via \proxystore{}'s Redis backend performs similar to without \proxystore{}, but larger objects see significant gains.
The Task Server, upon receiving a task from the Thinker, deserializes the task to determine the endpoint the task needs to be executed on and then serializes the task again to send to \funcx{}.
ProxyStore avoids additional deserialization and serialization of the input data because the data are replaced by a lightweight proxy.

A workflow using \colmena{} and \funcx{} for data movement will not be able to respond to or initiate tasks with low latency as data sizes increase.
Having an alternative method of object communication that can enable direct exchange of objects between the Thinker and workers is vital to avoid overloading the intermediate systems.

\subsubsection{\proxystore{} Backends}

\begin{figure*}[ht]
    \centering
    \includegraphics[trim=0.2cm 0.2cm 0.2cm 0.2cm, clip, width=\textwidth]{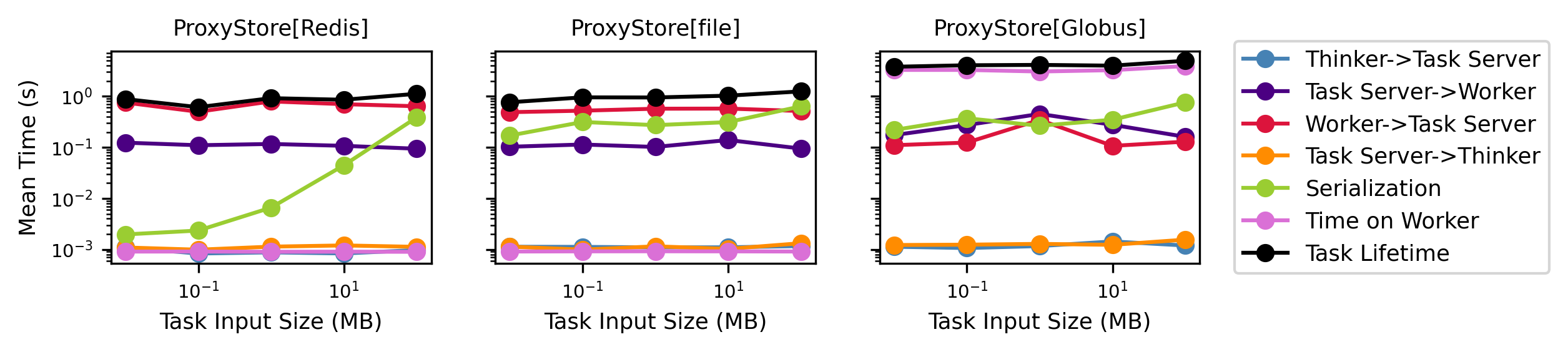}
    \caption{
    Mean times of components in the lifetime of a \colmena{} task. 
    The Redis, file system, and \globus{} \proxystore{} backends are used to proxy inputs, ranging in size from 10~kB to 100~MB, to no-op tasks.
    The Redis backend provides low latency while the file system backend performs well with large objects ($\sim$100~MB).
    Mean time of a task spent on a worker increases with \globus{} because the task must wait on the web-based data transfer to complete.
    }
    \label{fig:proxystore-input-size}
\end{figure*}

As discussed in \autoref{sec:proxystore-backends}, we employ multiple \proxystore{} backends to support a range of task characteristics and computing environments.
For example, Redis can be faster than file system I/O for certain object sizes, but opening ports for Redis may not be possible in all environments. 
Here we benchmark the ProxyStore backends to guide the deployment of our motivating applications.

Specifically, we measure components of a task's life cycle with three different \proxystore{} backends---Redis, file system, and \globus{}---across a range of task input sizes, from 10~kB to 100~MB.
We use the same no-op and no-output tasks, and the same Theta KNL node endpoint, as in the prior experiment.
The Redis and file system backends enable intra-site communication, and thus for experiments with those backends we place the Thinker and Task Server on the same Theta login node.
The \globus{} backend is used for inter-site communication, and so for Globus experiments we place the Thinker and Task Server on a login node of a cluster in the the University of Chicago's Research Computing Center.


Comparing the performance of Redis-backed and file system-backed \proxystore{} in \autoref{fig:proxystore-input-size}, we find the only significant differences to be in serialization time, which includes storing objects in the \proxystore{} backend.
Latencies are much lower when using Redis in the case of smaller objects, but are comparable across Redis and file system backends for larger objects.
The ``time on worker'' is consistent between the Redis and file system backends and across task input sizes.

The serialization time is similar between the file system and \globus{} backends.
Objects are still written to the shared file system prior to starting a \globus{} transfer operation.
Thus, the serialization performance in both cases is a reflection of the I/O performance of the file system.
The \globus{} backend does incur higher overheads in the time spent on the worker, because a proxy must wait for the \globus{} transfer of its target object to finish before the proxy can be resolved. 
This can take up to a few seconds, but the performance is constant with task input size (up to 100~MB), which indicates the communication bottleneck is due to the \globus{} web service latency rather than bandwidth limitations of the Thinker.

Overall, there are clear trade offs between backends for \proxystore{}.
A Redis backend is optimal for low latency access to small objects within a single site, whereas a shared file system backend is easier to use and provides good performance for large objects such as ML models.
\globus{} can enable multi-site workloads with low overheads when networking restrictions limit other options, and is competitive with Redis for dataset sizes beyond 10~MB.
All of these choices are accessible by changing a single line of code, enabling workflows to be easily optimized.

\subsection{Assessing System Performance in a Science Application}\label{results:design}

We next implement our science applications using our multi-resource workflow system
and ensure that communication overheads are not bottlenecks in system performance.
An effective steering system must minimize three forms of latency:
(a) between results completing and being available to the Thinker (reaction time),
(b) between results being received and new decisions being made (decision time),
and (c) between decisions made and new computations started (dispatch time).

\subsubsection{Reaction Time}

\begin{figure}
    \centering
    \includegraphics[width=\columnwidth,trim=0.4cm 0.5cm 0.35cm 0.3cm,clip]{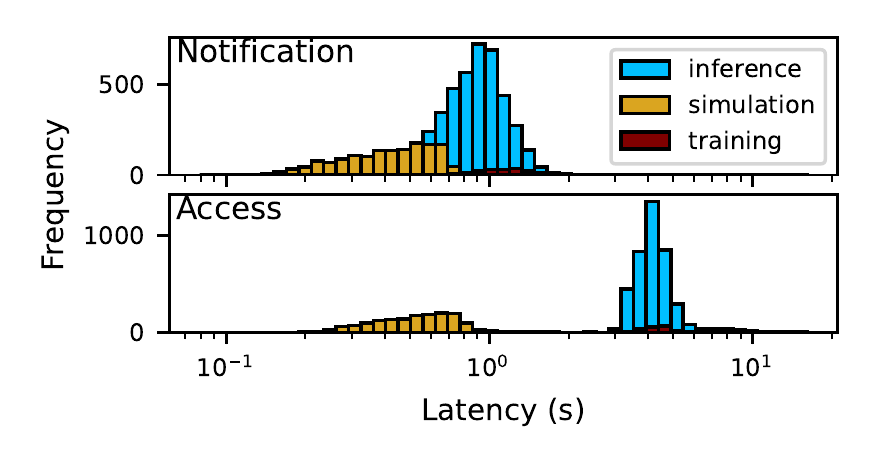}
    \caption{Result notification timings for Molecular Design application. Above: Time between when a tasks finishes computing and when the Thinker is notified that the computation has finished. Below: How long the Thinker waits to access the result data.}
    \label{fig:rxntime}
\end{figure}

Reaction time latency has two components: the time to notify the Thinker process of a task and the time to make the result data available. 
Our use of \proxystore{} means that the two processes occur separately.

Notification includes serializing the data, creating a proxy object, and communicating the result proxy to the Thinker.
As shown in \autoref{fig:rxntime}, notification for simulation tasks is faster (median \num{500}~ms) than inference or training as it does not require initializing a \globus{} transfer because the Thinker and simulation worker share a file system.
The inference and training tasks are limited by the latency to start a data transfer between resources, which requires an HTTPS request to \globus{} that takes an average of $\sim$\num{500}~ms.

Data transfer latency---the time taken to access a result---is only greater than 1~s when data are transferred between resources, as when moving between CPU and GPU machines for inference and training.
Transferring between resources requires a Globus transfer, which typically completes in 1--5~s, depending on data transfer node utilization and concurrent transfer limits per user.
Fusing multiple transfers into a single task would be a viable route to avoid the concurrent transfer limit and avoiding this issue.

\subsubsection{Decision Time}
Our steering policy is designed to minimize the points where decision making is time-critical.

The first latency-sensitive decision is to start a new simulation after another completes. 
This decision can be made rapidly because it does not require accessing the result data. 
The median time between a Thinker receiving a completed simulation and sending the next to the Task Server is only \num{5}~ms, which is negligible in comparison to Reaction Time.

Decisions based on model training and inference results involve reading the contents of a result, which requires deserializing the message and resolving any proxies.
Responding to model training and inference events each take a median of \num{4}~s; both times are primarily waiting for the data transfer to complete (as in \autoref{fig:rxntime}).
Thus the time to receive data from remote systems is, again, the most important factor in application performance.

\subsubsection{Dispatch Time}
The time for a chosen task to begin executing on a remote system, the \textit{dispatch time}, is the final type of latency we consider.
We hide the dispatch latency in many places with a backlog of queued tasks, but this latency cannot be hidden in three places: when we begin training models, when a batch of inference tasks are submitted after the first model completes training, and when a simulation worker completes.
In each case, we know at least one worker on a resource is available, and we want to provide this worker a new task as fast as possible.

Data transfer across multiple resources is the primary source of latency for inference and training tasks.
The median latencies for starting the first inference task of a batch, and a training task, are \num{3.8} and \num{2.5}~s, respectively; resolving the data proxy accounts for \num{3.6}~s (95\%) and \num{1.7}~s (67\%) of those times. 
Each of these latencies is small in terms of the average task times; overhead is less than 1\% of runtime for training tasks and less than 10\% inference tasks. 
The inference tasks benefit strongly from the ahead-of-time data transfer and caching provided by ProxyStore, with 12\% of inference proxies resolving in under \num{100}~ms ($<1$\% of the task time).

The latency of the simulation dispatch times is minimal.
The time is dominated by communicating the task request via \funcx{}, which is a median of \num{100}~ms.
The dispatch time is less than 1\% of the total task runtime and, consequently, not our main target for improvement.

\subsubsection{Assessment} The latency of responding to a completed calculation across multiple resources is typically below \num{1}~s but can be as high as \num{10}~s under worst-case conditions.
As illustrated in the previous sections, \colmena{} is alerted of a new result within \num{100}~ms and acting on the result can take up to \num{3}~s if data must be transferred from a remote system.
Starting a new task on a new system is a minimum of \num{100}~ms and increases to several seconds if data must be transferred between sites.

\subsection{Comparing System Performance}\label{results:compare}

We study the performance differences between workflow systems (see \autoref{sec:workflow}) with two metrics: scientific performance and responsiveness of the workflow system. 
 
\subsubsection{Molecular Design}

\begin{figure}
    \centering
    \includegraphics[trim=3mm 0.1cm 3mm 0cm, clip,width=\columnwidth]{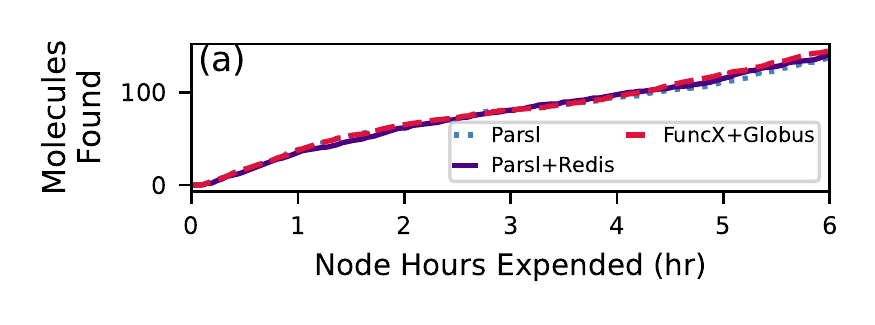}
    \includegraphics[trim=3mm 0.4cm 3mm 0.3cm, clip,width=\columnwidth]{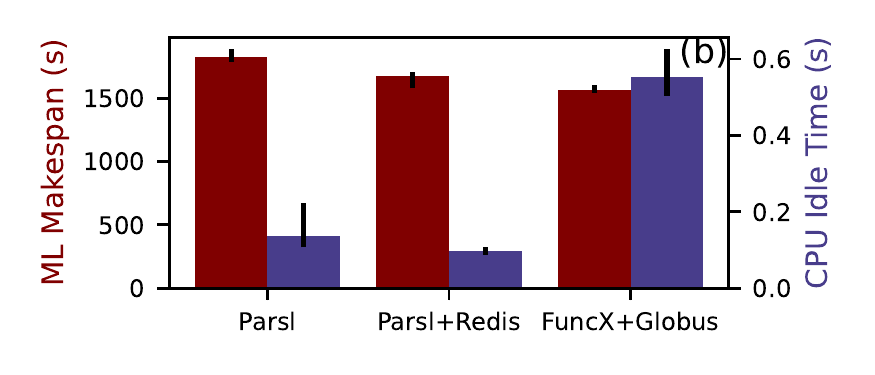}
    \caption{Comparison of different implementations of the multi-site active learning pipeline. An implementation using only Parsl, Parsl using \proxystore{} with Redis, and \funcx{} 
    using \proxystore{} with 
    \globus{}. (a) Number of top-performing molecules found as a function of simulation time expended.(b) Median time required to reorder a task queue (ML Makespan) and median idle time for CPU workers between simulation tasks. Error bars represent the 40th and 60th percentiles. }
    \label{fig:parslcomparison}
\end{figure}

We compared scientific outcomes by measuring how many molecules with IP $>$ 14 were found after 6 node hours of compute and averaged the value over three runs of each implementation.
Our best effort with open ports, Parsl+Redis, has equivalent outcomes to \funcx{}+\globus{} (140.3 vs.\ 145.0). 
The results are within the margin of error as the Parsl+Redis version varied between 129 and 149 suitable molecules found over 3 runs with different random seeds.
We therefore conclude that direct connections between computing providers are an unnecessary complication to deploying our multi-resource application.

The key responsiveness needed by our application is the time to update the task list given new data and rate at which new simulations are evaluated.
We define the time to update the task list as the time from when the steering policy requests models to be retrained to when the results from all inference tasks are used to reprioritize the task queue.
As shown in \autoref{fig:parslcomparison}a, the \funcx{} implementation completes the ML tasks faster (in 1565~s, on average) than native Parsl (1828~s) and is faster than Parsl+\proxystore{} backed by Redis (1676~s). 
There is a clear advantage to using pass-by-reference, with both \proxystore{}-backed applications outperforming a baseline Parsl, and we find that transferring with Globus yields improved application performance given the large data requirements of inference tasks.

Our other key latency metric is CPU utilization.
CPU nodes are idle between one simulation completing and the next starting, which requires two fast operations: notifying the Thinker of a completion and dispatching the next task.
We find the average idle time between tasks around 500~ms for FuncX and around 100~ms for Parsl with a Redis \proxystore{}.
In both cases, these latencies are small enough to achieve overall CPU utilization of above 99\%.
Utilization can be improved even further improved by submitting at least one more simulation task to execute than there are CPU workers available.

\subsubsection{Surrogate Finetuning}
\begin{figure}
    \centering
    \includegraphics[trim=4mm 5mm 4mm 3mm, clip,width=\columnwidth]{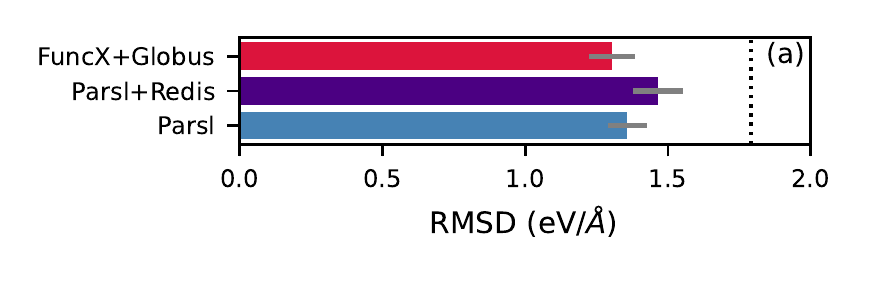}
    \includegraphics[trim=3mm 4mm 3mm 1mm, clip,width=\columnwidth]{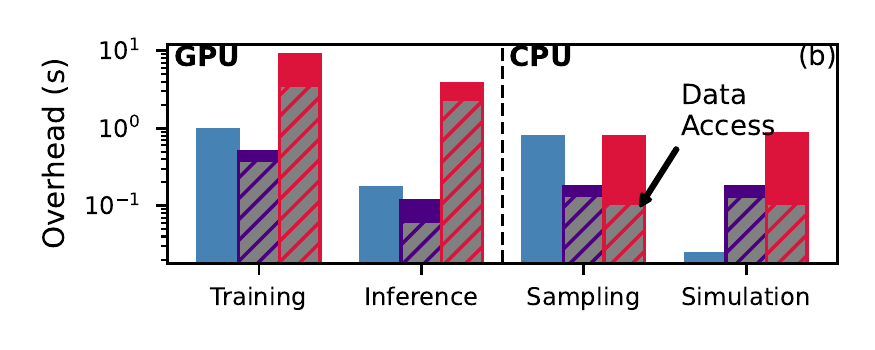}
    \caption{Comparison of (a) Root Mean Squared Deviation (RMSD) in predicted forces on a test set and (b) median overheads per task type for three different workflow systems. (a) Error bars are the standard error of the mean over three tests. The vertical dashed line is the error before fine-tuning. (b) The time spent waiting for result data is shown in gray for the systems where \proxystore{} is used to transfer data separately from task instructions (Redis store with Parsl, Globus store with FuncX).}
    \label{fig:surrogate}
\end{figure}

The surrogate models produced using our cloud-managed workflows are indistinguishable from those produced using a conventional workflow solution.
The Root Mean Squared Deviation (RMSD) of the forces predicted using the models from the FuncX solution are $1.30\pm0.08$ $\mathrm{eV/\text{\AA}}$, compared to $1.47\pm0.09$ $\mathrm{eV/\text{\AA}}$ for Parsl with ProxyStore and $1.36\pm0.07$ $\mathrm{eV/\text{\AA}}$ without (\autoref{fig:surrogate}a). 
Run-to-run variations are larger than variation between the applications. 

While the three approaches show comparable scientific performance, the task overhead when using FuncX and Globus is clearly larger.
We measure this overhead as the time between when a task was created and when the result was read that is not the task running.
When tasks are run on remote GPUs, overhead is dominated by the time to transfer data with Globus (\autoref{fig:surrogate}b).
The second largest component is the data transfer time, which is roughly 2~s per direction and consistent with the results of synthetic experiments (\autoref{fig:proxystore-input-size}).
Transferring to remote sites via Redis is faster, but requires configuring a tunnel between resources.

CPU task overheads are dominated by the time to notify the Thinker of task completion for FuncX and data transfer for Parsl, regardless of task data size.
In contrast, the overhead for Parsl without pass-by-reference is strongly dependent on data size; 820~ms for sampling tasks (3~MB) and 20~ms for simulation tasks (20~kB).
The data transfer times for Parsl with pass-by-reference are consistent for these two task types at 200 and 170~ms for the sampling and simulation, respectively.
We can see the benefit of pass-by-reference for larger messages but also that our application could be accelerated by avoiding the overhead of proxying small messages.

\subsection{Recommendations}

Our experience leads to several recommendations.
\begin{itemize}
    \item Use pass-by-reference and intelligent steering policies to achieve 10x reductions in application latency.
    \item Transmit data between sites directly for data larger than 10~kB. If messages are smaller than 100~MB and direct connection between resources is feasible, Redis is ideal, otherwise Globus is the best choice.
    \item Employ pass-by-reference with a conventional workflow system (e.g., Parsl) if data are larger than 10~kB, especially if data are reused between tasks.
\end{itemize}

\section{Conclusions}

We have presented experiences deploying two scientific applications across multiple heterogeneous resources.
The applications share a need to run simulation and AI tasks on separate resources, but differ in the amount of data transfer by an order of magnitude and in the frequency of AI tasks being required.
We employed a combination of FuncX and Globus Transfer to pass task instructions and data between resources,
\proxystore{} to allow data transfer via Globus without modifying application code,
and \colmena{} to express steering policies that hide data transfer latencies.
We show that our implementation achieves scientific outputs that are indistinguishable from those produced by implementations that require direct connections between resources.
We hope our demonstration that cloud services simplify deployment and ensure security of AI-Integrated workflows without reducing performance will lower the barrier to more applications being deploy across computational sciences.




\section*{Acknowledgments}
We thank K.~Hermann, H.~Sprueill, J.~Bilbrey, and S.~Xantheas for help with the surrogate fine-tuning application.
LW, GS, GP, RC, SC, RT, and IF acknowledge support by the ExaLearn Co-design Center of Exascale Computing Project (17-SC-20-SC), a collaborative effort of the U.S. Department of Energy Office of Science and the National Nuclear Security Administration, to develop Colmena and evaluate its performance on HPC.
YB and KC were supported to integrate Parsl with Colmena by NSF Grant 1550588 and the ExaWorks Project within the Exascale Computing Project.
GP, VHS, and KC were supported to develop ProxyStore by NSF Grant 2004894.
This research used resources of the Argonne Leadership Computing Facility (ALCF), a DOE Office of Science User Facility supported under Contract DE-AC02-06CH11357, including via the ALCF Data Science Program. It also used resources provided by the University of Chicago's Research Computing Center.

\balance

\bibliographystyle{IEEEtran}
\bibliography{IEEEabrv, refs}

\begin{thebibliography}{10}
\providecommand{\url}[1]{#1}
\csname url@samestyle\endcsname
\providecommand{\newblock}{\relax}
\providecommand{\bibinfo}[2]{#2}
\providecommand{\BIBentrySTDinterwordspacing}{\spaceskip=0pt\relax}
\providecommand{\BIBentryALTinterwordstretchfactor}{4}
\providecommand{\BIBentryALTinterwordspacing}{\spaceskip=\fontdimen2\font plus
\BIBentryALTinterwordstretchfactor\fontdimen3\font minus
  \fontdimen4\font\relax}
\providecommand{\BIBforeignlanguage}[2]{{%
\expandafter\ifx\csname l@#1\endcsname\relax
\typeout{** WARNING: IEEEtran.bst: No hyphenation pattern has been}%
\typeout{** loaded for the language `#1'. Using the pattern for}%
\typeout{** the default language instead.}%
\else
\language=\csname l@#1\endcsname
\fi
#2}}
\providecommand{\BIBdecl}{\relax}
\BIBdecl

\bibitem{vescovi2022linking}
R.~Vescovi \emph{et~al.}, ``Linking scientific instruments and computation:
  Patterns, technologies, and experiences,'' \emph{Patterns}, vol.~3, no.~10,
  2022.

\bibitem{brace2021stream}
A.~Brace \emph{et~al.}, ``{Stream-AI-MD}: Streaming {AI}-driven adaptive
  molecular simulations for heterogeneous computing platforms,'' in
  \emph{Platform for Advanced Scientific Computing Conference}, 2021, pp.
  1--13.

\bibitem{smarr1992metacomputing}
L.~Smarr and C.~E. Catlett, ``Metacomputing,'' \emph{Communications of the
  ACM}, vol.~35, no.~6, pp. 44--52, 1992.

\bibitem{brunett1998application}
S.~Brunett \emph{et~al.}, ``Application experiences with the {G}lobus
  toolkit,'' in \emph{7th Intl Symp on High Performance Dist Computing}, 1998,
  pp. 81--88.

\bibitem{chard2014globus}
K.~Chard, S.~Tuecke, and I.~Foster, ``Efficient and secure transfer,
  synchronization, and sharing of big data,'' \emph{IEEE Cloud Computing},
  vol.~1, no.~3, pp. 46--55, 2014.

\bibitem{chard20funcx}
R.~Chard \emph{et~al.}, ``func{X}: A federated function serving fabric for
  science,'' in \emph{29th Intl Symp on High-Performance Parallel Dist
  Computing}, 2020.

\bibitem{babuji19parsl}
Y.~Babuji \emph{et~al.}, ``Parsl: Pervasive parallel programming in {P}ython,''
  in \emph{Intl Symp on High-Performance Parallel and Dist Computing}, 2019.

\bibitem{feller2007gt4}
M.~Feller, I.~Foster, and S.~Martin, ``{GT4 GRAM}: A functionality and
  performance study,'' in \emph{TeraGrid Conference}, 2007.

\bibitem{cholia10newt}
S.~Cholia, D.~Skinner, and J.~Boverhof, ``Newt: A restful service for building
  high performance computing web applications,'' in \emph{Gateway Computing
  Environments Workshop}, 2010, pp. 1--11.

\bibitem{stubbs21tapis}
J.~Stubbs \emph{et~al.}, ``Tapis: An {API} platform for reproducible,
  distributed computational research,'' in \emph{Advances in Information and
  Communication}, K.~Arai, Ed., 2021, pp. 878--900.

\bibitem{deelman19pegasus}
E.~Deelman \emph{et~al.}, ``The evolution of the {P}egasus workflow management
  software,'' \emph{Computing in Science \& Engineering}, vol.~21, no.~4, 2019.

\bibitem{rocklin2015dask}
M.~Rocklin, ``Dask: Parallel computation with blocked algorithms and task
  scheduling,'' in \emph{14th Python in Science Conference}, 2015.

\bibitem{zhao07swift}
Y.~Zhao \emph{et~al.}, ``Swift: Fast, reliable, loosely coupled parallel
  computation,'' in \emph{IEEE Congress on Services}, 2007, pp. 199--206.

\bibitem{afgan22galaxy}
E.~Afgan \emph{et~al.}, ``The {G}alaxy platform for accessible, reproducible
  and collaborative biomedical analyses: 2022 update,'' \emph{Nucleic Acids
  Research}, vol.~50, no.~W1, pp. W345--W351, Apr. 2022.

\bibitem{bala19radical}
V.~Balasubramanian \emph{et~al.}, ``{RADICAL-Cybertools}: Middleware building
  blocks for scalable science,'' 2019.

\bibitem{amazonlambda}
{Amazon Lambda}. \url{https://aws.amazon.com/lambda}. Accessed Jan 2023.

\bibitem{googlecloudfunctions}
{Google Cloud Functions}. \url{https://cloud.google.com/functions/}. Accessed
  Jan 2023.

\bibitem{azureFunctions}
{Azure Functions}. \url{https://azure.microsoft.com/en-us/services/functions/}.
  Accessed Jan 2023.

\bibitem{foster2017cloud}
I.~Foster and D.~B. Gannon, \emph{Cloud Computing for Science and
  Engineering}.\hskip 1em plus 0.5em minus 0.4em\relax MIT Press, 2017.

\bibitem{spillner2017faaster}
J.~Spillner, C.~Mateos, and D.~A. Monge, ``Faaster, better, cheaper: The
  prospect of serverless scientific computing and {HPC},'' in \emph{Latin
  American High Performance Computing Conference}, 2017, pp. 154--168.

\bibitem{malawski2016towards}
M.~Malawski, ``Towards serverless execution of scientific
  workflows--{HyperFlow} case ssudy,'' in \emph{Workshop on Workflows in
  Support of Large-Scale Science}, 2016, pp. 25--33.

\bibitem{fox2017conceptualizing}
G.~Fox and S.~Jha, ``Conceptualizing a computing platform for science beyond
  2020,'' in \emph{10th Intl Conference on Cloud Computing}, 2017.

\bibitem{kiar2019serverless}
G.~Kiar \emph{et~al.}, ``A serverless tool for platform agnostic computational
  experiment management,'' \emph{Frontiers in Neuroinformatics}, vol.~13, 2019.

\bibitem{openwhisk}
{Apache OpenWhisk}. \url{http://openwhisk.apache.org/}. Accessed Jan 2023.

\bibitem{Fn}
Fn project. \url{https://fnproject.io}. Accessed Jan 2023.

\bibitem{knix}
{KNIX MicroFunctions}. \url{https://github.com/knix-microfunctions/knix}.
  Accessed Jan 2023.

\bibitem{garcia2020abaco}
C.~Garcia \emph{et~al.}, ``Abaco--a modern platform for highthroughput parallel
  scientific computations,'' in \emph{Intl Workshop on Science Gateways}, 2020.

\bibitem{ghaemi2020chainfaas}
S.~Ghaemi, H.~Khazaei, and P.~Musilek, ``Chain{FaaS}: An open blockchain-based
  serverless platform,'' \emph{IEEE Access}, vol.~8, 2020.

\bibitem{ciavotta2021dfaas}
M.~Ciavotta \emph{et~al.}, ``{DFaaS}: Decentralized function-as-a-service for
  federated edge computing,'' in \emph{10th Intl Conf Cloud Networking}, 2021.

\bibitem{carriero1994linda}
N.~J. Carriero \emph{et~al.}, ``The {L}inda alternative to message-passing
  systems,'' \emph{Parallel Computing}, vol.~20, no.~4, pp. 633--655, 1994.

\bibitem{dataspaces2017aktas}
M.~F. Aktas \emph{et~al.}, ``{WA-Dataspaces}: Exploring the data staging
  abstractions for wide-area distributed scientific workflows,'' in \emph{46th
  Intl Conference on Parallel Processing}, 2017, pp. 251--260.

\bibitem{soumagne2013mercury}
J.~Soumagne \emph{et~al.}, ``Mercury: Enabling remote procedure call for
  high-performance computing,'' in \emph{Intl Conf Cluster Computing}, 2013,
  pp. 1--8.

\bibitem{ross2020mochi}
R.~Ross \emph{et~al.}, ``Mochi: Composing data services for high-performance
  computing environments,'' \emph{J Computer Science and Technology}, 2020.

\bibitem{aktas2017wa}
M.~F. Aktas \emph{et~al.}, ``{WA-DataSpaces}: Exploring the data staging
  abstractions for wide-area distributed scientific workflows,'' in \emph{46th
  Intl Conference on Parallel Processing}, 2017, pp. 251--260.

\bibitem{data2013sciencedmz}
E.~Dart \emph{et~al.}, ``{The Science DMZ}: A network design pattern for
  data-intensive science,'' ser. SC'13, 2013.

\bibitem{chung2022scistream}
J.~Chung \emph{et~al.}, ``Sci{S}tream: Architecture and toolkit for data
  streaming between federated science instruments,'' in \emph{31st Intl Symp on
  High-Performance Parallel and Distributed Computing}, 2022, p. 185–198.

\bibitem{jha2022AIHPC}
\BIBentryALTinterwordspacing
S.~Jha, V.~R. Pascuzzi, and M.~Turilli, ``{AI}-coupled {HPC} workflows,'' 2022.
  [Online]. Available: \url{https://arxiv.org/abs/2208.11745}
\BIBentrySTDinterwordspacing

\bibitem{ejarque2022aiworkflows}
J.~Ejarque \emph{et~al.}, ``Enabling dynamic and intelligent workflows for
  {HPC}, data analytics, and {AI} convergence,'' \emph{Future Generation
  Computer Systems}, vol. 134, pp. 414--429, Sep. 2022.

\bibitem{atlas2022AtlFast3}
{ATLAS Collaboration}, ``Atl{F}ast3: The next generation of fast simulation in
  {ATLAS},'' \emph{Computing and Software for Big Science}, vol.~6, p.~7, 2022.

\bibitem{dunn2019rocketsled}
A.~Dunn, J.~Brenneck, and A.~Jain, ``Rocketsled: A software library for
  optimizing high-throughput computational searches,'' \emph{J Physics:
  Materials}, vol.~2, no.~3, p. 034002, Apr. 2019.

\bibitem{montoya2020camd}
J.~H. Montoya \emph{et~al.}, ``Autonomous intelligent agents for accelerated
  materials discovery,'' \emph{Chemical Science}, vol.~11, no.~32, 2020.

\bibitem{lee2019deepdrivemd}
H.~Lee \emph{et~al.}, ``{DeepDriveMD}: Deep-learning driven adaptive molecular
  simulations for protein folding,'' in \emph{3rd Workshop on Deep Learning on
  Supercomputers}, Nov. 2019.

\bibitem{wozniak2018supervisor}
J.~M. Wozniak \emph{et~al.}, ``{CANDLE}/{S}upervisor: A workflow framework for
  machine learning applied to cancer research,'' \emph{{BMC} Bioinformatics},
  vol.~19, no. S18, Dec. 2018.

\bibitem{balaprakash2018deephyper}
P.~Balaprakash \emph{et~al.}, ``{DeepHyper}: Asynchronous hyperparameter search
  for deep neural networks,'' in \emph{25th Intl Conference on HPC}, 2018.

\bibitem{libEnsemble}
S.~Hudson \emph{et~al.}, ``{libEnsemble} users manual,'' Argonne National
  Laboratory, Tech. Rep. Revision 0.7.1, 2020.

\bibitem{mortiz2018ray}
P.~Moritz \emph{et~al.}, ``Ray: A distributed framework for emerging {AI}
  applications,'' in \emph{13th {USENIX} OSDI}, 2018, pp. 561--577.

\bibitem{yildiz2021decaf}
O.~Yildiz \emph{et~al.}, ``Dynamic heterogeneous task specification and
  execution for in situ workflows,'' in \emph{Workshop on Workflows in Support
  of Large-Scale Science}, Nov. 2021.

\bibitem{ward2023data}
\BIBentryALTinterwordspacing
L.~Ward \emph{et~al.}, ``Dataset for cloud services enable efficient
  {AI}-guided simulation workflows across heterogeneous resources,'' 2023.
  [Online]. Available:
  \url{https://petreldata.net/mdf/detail/multiresource_ai_v2.1}
\BIBentrySTDinterwordspacing

\bibitem{polykovskiy2020moses}
D.~Polykovskiy \emph{et~al.}, ``{M}olecular {S}ets ({MOSES}): A benchmarking
  platform for molecular generation models,'' \emph{Frontiers in Pharmacology},
  2020.

\bibitem{babuji2020covdata}
Y.~Babuji \emph{et~al.}, ``Targeting {SARS-CoV-2 with AI- and HPC}-enabled lead
  generation: A first data release,'' 2020, arXiv:2006.02431.

\bibitem{cohn1996active}
D.~A. Cohn, Z.~Ghahramani, and M.~I. Jordan, ``Active learning with statistical
  models,'' \emph{J Artificial Intelligence Research}, vol.~4, 1996.

\bibitem{li2020recent}
M.~Li \emph{et~al.}, ``Recent advancements in rational design of non-aqueous
  organic redox flow batteries,'' \emph{Sustainable Energy \& Fuels}, vol.~4,
  no.~9, pp. 4370--4389, 2020.

\bibitem{landrum2006rdkit}
G.~Landrum, ``{RDKit}: Open-source cheminformatics,''
  \url{https://www.rdkit.org}. Visited May 1, 2021.

\bibitem{leeping2016geometric}
L.-P. Wang and C.~Song, ``Geometry optimization made simple with translation
  and rotation coordinates,'' \emph{J Chemical Physics}, vol. 144, no.~21, p.
  214108, 2016.

\bibitem{smith2020qcarchive}
D.~G.~A. Smith \emph{et~al.}, ``The {MolSSI} {QCA}rchive project: An
  open-source platform to compute, organize, and share quantum chemistry
  data,'' \emph{{WIREs} Computational Molecular Science}, vol.~11, no.~2, 2020.

\bibitem{bannwarth2020xtb}
C.~Bannwarth \emph{et~al.}, ``Extended tight‐binding quantum chemistry
  methods,'' \emph{{WIREs} Comput{.} Mol{.} Sci{.}}, vol.~11, p. e01493, 2020.

\bibitem{stjohn2019mpnnpolymers}
P.~{St. John} \emph{et~al.}, ``Message-passing neural networks for
  high-throughput polymer screening,'' \emph{J Chemical Physics}, vol. 150,
  no.~23, 2019.

\bibitem{fanourgakis2006ttm}
G.~S. Fanourgakis and S.~S. Xantheas, ``The flexible, polarizable, thole-type
  interaction potential for water ({TTM}2-f) revisited,'' \emph{J Physical
  Chemistry A}, vol. 110, no.~11, pp. 4100--4106, Mar. 2006.

\bibitem{choudhury2020hydronet}
S.~Choudhury \emph{et~al.}, ``Hydro{N}et: Benchmark tasks for preserving
  intermolecular interactions and structural motifs in predictive and
  generative models for molecular data,'' \emph{Machine Learning and the
  Physical Sciences Workshop at NeurIPS}, 2020.

\bibitem{schutt2018schnet}
K.~T. Sch\"{u}tt \emph{et~al.}, ``{SchNet} {\textendash} a deep learning
  architecture for molecules and materials,'' \emph{J Chemical Physics}, vol.
  148, no.~24, p. 241722, 2018.

\bibitem{smith2020psi4}
D.~Smith \emph{et~al.}, ``Psi4 1.4: Open-source software for high-throughput
  quantum chemistry,'' \emph{J Chemical Physics}, vol. 152, no.~18, 2020.

\bibitem{larsen2017ase}
A.~H. Larsen \emph{et~al.}, ``The atomic simulation environment{\textemdash}a
  {Python} library for working with atoms,'' \emph{J Physics: Condensed
  Matter}, vol.~29, no.~27, p. 273002, Jun. 2017.

\bibitem{tuecke2016globus}
S.~Tuecke \emph{et~al.}, ``Globus {A}uth: A research identity and access
  management platform,'' in \emph{12th Intl Conf e-Science}, 2016, pp.
  203--212.

\end{thebibliography}

\end{document}